\documentclass[a4paper,twoside]{article}

\usepackage{graphicx,subfigure,amssymb,amsmath,natbib}

\title{Mathematic principles underlying genetic structures}
\author{Matthew J. Berryman}
\date{}
\begin{document}
\maketitle

\begin{abstract}
Many people are familiar with the physico-chemical properties of gene sequences. In this paper I present a mathematical perspective: how do mathematical principles such as information theory, coding theory, and combinatorics influence the beginnings of life and the formation of the genetic codes we observe today? What constraints on possible life forms are imposed by information-theoretical concepts? Further, I detail how mathematical principles can help us to analyse the genetic sequences we observe in the world today.
\end{abstract}

\section{Introduction}
\begin{figure}[htbp]
\centering
\includegraphics[width=7cm]{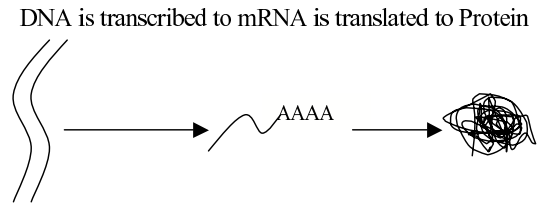}
\caption{The central dogma of molecular biology states that ``DNA is transcribed into messenger RNA, which is then translated into protein.'' This diagram also shows DNA replication, which is done with the aid of a number of proteins. At the mRNA stage, introns are spliced out from the sequence, leaving only the protein coding exons. This dogma is of course vastly simplified, for example there is added complexity through splicing, RNA-only genes, RNA-RNA interactions, prions, and other details~\cite{NGDogma,DarwinGenome}. But in its essential form this does describe the flow of information in a cell.}
\label{DogmaFigure}
\end{figure}

Genetics is concerned with the physical characteristics of organisms that are passed on
from one organism to another through the use of deoxyribonucleic acid (DNA), consisting of
a sequence of nucleotides. The nucleotides are the chemical bases adenosine, thymine, cytosine and guanine that are denoted using the alphabet $\left\{\mathrm{A,T,C,G}\right\}$. Those on one strand are paired in a complementary fashion with those on the other strand, where adenosine matches with thymine, and guanine with cytosine. Groups
of three bases are called codons, and these encode the twenty amino acids that combine to
form proteins, the building blocks of life. In a nutshell, the central dogma of molecular biology states that 
``DNA makes RNA makes protein''. This is encapsulated in Figure~\ref{DogmaFigure}. The DNA is transcribed into complementary messenger ribonucleic acid (mRNA). In RNAs, the alphabet is 
$\left\{\mathrm{A,T,U,G}\right\}$ where uracil plays the same role that thymine does in DNA, as it pairs with guanine.
Sections of the mRNA that do not code for proteins are removed, and a ``poly-A tail''---a sequence composed entirely of adenosine bases---is added to (chemically) stabilise the sequence. The mRNA then acts as a template for protein synthesis. Transfer RNAs (tRNAs) bind to an amino acid on one end, and a complimentary set of three bases on the mRNA template.  A 1D sequence of amino acids forms and is then detached from the tRNAs and folds into a 3D structure. This sometimes occurs by itself and sometimes with the aid of other proteins, either immediately or at a later date in the life of the cell. 

There are several key areas in which mathematical principles underlie, influence, and can provide information about genetic structures. The key questions that these principles can help answer are
\begin{itemize}
\item Why do we have four bases, a triplet coding and twenty amino acids?
\item Why do we observe the particular assignment of triplets to amino acids that we do?
\item How do new gene sequences arise, and how do they spread in a population?
\item How can we analyse the sequences that arise?
\end{itemize}
Some mathematically-based answers are discussed in the remainder of this paper. 

\section{Number of bases and amino acids}
\begin{figure}[hptb]
\centering
\includegraphics[width=4in]{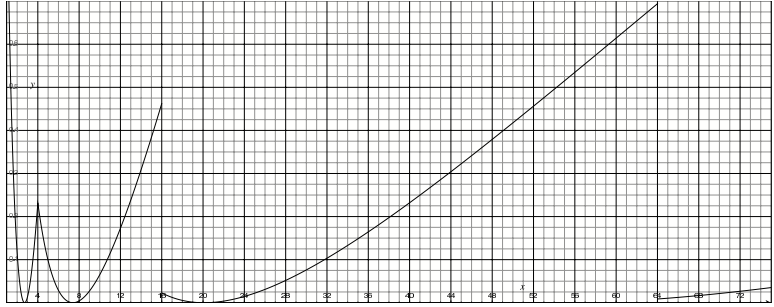}
\caption{Plot of the distance between the actual hardware and optimal hardware, on the vertical axis, for a different number of amino acids. }
\label{fig:Delta}
\end{figure}

The following is a summary of the work of Soto and Toh\'{a}~\cite{DNA_Hardware}, who took a mathematical approach to the question of why four bases, a triplet coding, and 20 amino acids are used, based on the assumption that nature will, over evolutionary time, find a solution to the problem that minimises the amount of cell machinery. It also assumes that the machinery is not unlike that used by computer memory chips to decode~\cite{DNA_Hardware,MemOptimalRadix}. This is not a bad assumption, but leaves out chemical tricks that that the tRNAs can use. I also use the fact that optimal solutions, since they have an advantage in evolutionary terms, spread in a population as I explore later. The main argument of Soto and Toh\'{a} is as follows:

Firstly, they define the maximum number of amino acids as 
\begin{equation}
N=p^n,
\label{eqn:N}
\end{equation}
where $p$ is the number of possible bases (symbols of length 1) and $n$ the number of positions. For example, the amino acid codings used in all living things has $p=4$ bases and $n=3$ positions, a triplet code. This gives a total of $N=64$ possible amino acids. 
For the assumptions above, it turns out the amount of ``hardware'', or cell machinery, is proportional to the number of bases times the number of positions, which can be written as
\begin{equation}
\mathrm{hardware}\propto p\times n,
\label{eqn:hardware}
\end{equation}
where $p$ and $n$ are as defined above. 
It also turns out that to minimise the amount of hardware, one can write this number of amino acids as,
\begin{equation}
A=e^x,
\label{eqn:A}
\end{equation}
where $e=2.718...$ is the base of the natural logarithm, and describes many growth and decay processes that occur in the natural world, and $x$ is the number of positions. So we need to have the number of bases close to $e$, thus optimising the number of positions for a given $N$ by setting
\begin{equation}
x=\ln N.
\label{x}
\end{equation}
Then we can find a semi-optimal $N$ by,
\begin{equation}
N=(N^{1/y})^y,
\label{eqn:NsemiOpt}
\end{equation}
where $y$ is the actual number of positions used, resulting in a degeneracy,
\begin{equation}
D=\left(\frac{p}{b}\right)^y,
\label{D}
\end{equation}
where $p$ is the number of actual bases used and $b$ is the minimum (integer) possible.
Then the actual amount of hardware used is
\begin{equation}
B=b\times y=yN^{1/y},
\label{B}
\end{equation}
and we write the difference between this amount of hardware, and the optimal, $A=e^x$, as
\begin{equation}
\Delta =B-A\geq 0,
\label{eqn:Delta}
\end{equation}
where $\Delta$ is the difference in ``hardware'' between the actual and optimal solution, and this is always greater than zero as we can approach but never achieve the minimal amount of ``hardware'' (since this would require a non-integral number of bases.
If we set the derivative, or rate of change of $\Delta$,
\begin{equation}
\frac{d\Delta}{dN}=\frac{d}{dN}\left(yN^{1/y}-e\ln N\right),
\label{eqn:dDelta:dN}
\end{equation}
to zero, this allows us to find the optimal solution for the number of amino acids for fixed number of base positions.
A graph of $\Delta$ is shown in Figure~\ref{fig:Delta}, showing the minima for one, two and three positions occurring at three, seven, and 20. This assumes four bases are used The actual minima, and for the best possible choice of number of bases, are shown in Table~\ref{table:minima}, again, indicating 20 amino acids is the optimal number. 
\begin{table}[hptb]
\centering
\caption{This table shows the optimal number of amino acids for 1-4 base positions, and the corresponding (minimal) difference between the actual coding and the theoretical minimum $A=e^x$.}
\begin{tabular}{|c|c|c|c|}\hline
$y$ (number of positions) & $N$ (number of amino acids) & rounded $N$ & $\Delta$\\\hline\hline
1 & 2.718 & 3 & 0.0136\\\hline
2 & 7.389 & 7 & 0.0019\\\hline
3 & 20.086 & 20 & $8.2\times10^{-6}$\\\hline
5 & 54.598 & 55 & $18\times10^{-6}$\\\hline
\end{tabular}
\label{table:minima}
\end{table}

Having established that 20 amino acids is best, we can then turn to the problem of why four bases are used. It is shown above that the best we can do is to get as close to $e$ bases as possible, by choosing three amino acids. Two bases would require more positions (and more hardware) than three, or four. There are two main reasons why four bases is actually a better choice than three, however:
\begin{enumerate}
\item Four bases allows a complimentary pairing, for accurate, fast and efficient replication of genetic material.
\item On the hypothesis that there is a precursor genetic code, using fewer positions and coding less than 20 amino acids~\cite{patel-2005-233}, then this evolutionary pathway is actually easier (in terms of more efficient in hardware) if four bases are used. Table~\ref{table:threeVSfour} shows the corresponding lower values of $\Delta$ (normally lower for three, but not always as this Table shows).
\end{enumerate}
\begin{table}[hptb]
\centering
\caption{The following values of $\Delta$, lower for four bases than three, indicate the pathway to 20 amino acids and four bases.}
\begin{tabular}{|c|c|c|}\hline
$y$ (number of positions) & $\Delta$ for 3 bases & $\Delta$ for 4 bases\\\hline\hline
4 &	0.2317 &	0.2317\\\hline
10 &	0.2042 &	0.0655\\\hline
11 &	0.1538 &	0.1151\\\hline
\end{tabular}
\label{table:threeVSfour}
\end{table}

\section{Gray coding}
Mitochondria are organelles that live inside each cell, and provide the cell with energy. They have their own genetic code, independent of the normal set of 23 pairs of chromosomes that reside in the cell nucleus. They also use a different way of coding for the amino acids. The mitochondrial code for vertebrates is shown in Table~\ref{fig:mitochondrialcode}. This differs slightly from the normal code used in nuclear DNA in that the ``wobble rule'' is exact. This means, that, for some particular choices of the first two bases, if we change the third base, we end up with the same amino acid in the code. This is important for reducing the cell machinery needed for decoding, as discussed above, but since mutations are occuring, we would like to have the result that a single mutation in a triple  (which is a number of times more likely than a double mutation in a triplet) results in either no change in the amino acid, or to a very chemically similar amino acids. 
\begin{table}[htbp]
\caption{Table of the RNA triplets and the corresponding (abbreviated) amino acid. The ``wobble rule'' can be seen as when the third base changes, we usually stay on the same amino acid, or in some cases move to a chemically related one.}
\centering
\includegraphics[width=7cm]{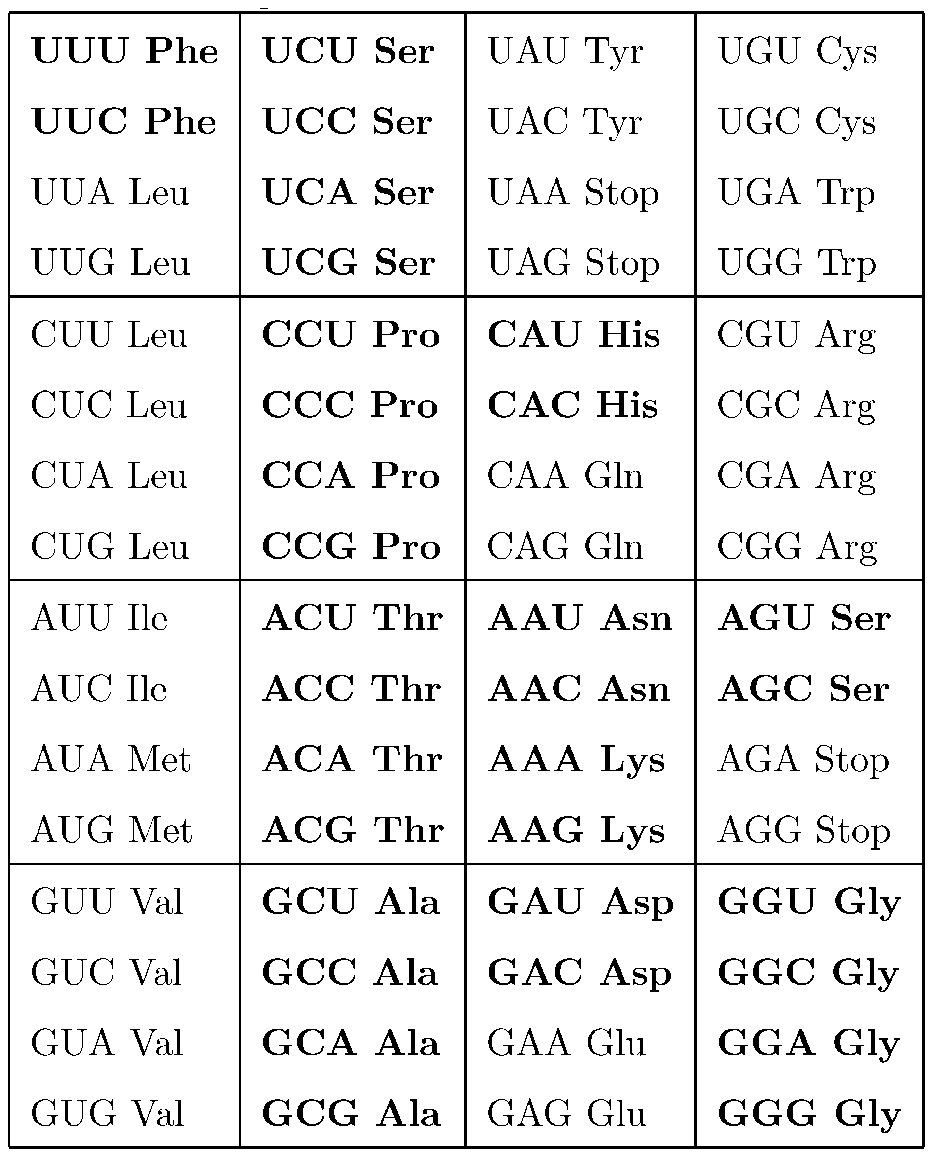}
\label{fig:mitochondrialcode}
\end{table}

We then have the problem of assigning the 20 amino acids (including that for the amino acid methionine, which doubles as a start codon) plus a stop codon to the set of 64 codons, such that a change by one base in the codon, results in minimal change in the amino acid, even in the non-``wobble rule'' (3rd base) position. The problem of doing this has shown by \citep{GeneticGrayCode} to be an equivalent problem to the travelling salesman problem---that is, solving one problem gives you a solution to the other problem. The travelling salesman problem is a very famous problem in the areas of discrete combinatorics---how to solve problems of arrangement of items---and in the theory of computational complexity, where we are interested in how long it takes to solve a difficult problem. The travelling salesman problem is how to visit a group of cities, visiting each city once, and going back to the starting city, all for minimum cost. It also turns out that this is an equivalent problem to the Towers of Hanoi game, in which one attempts to shift a set of discs stacked from largest up to smallest from one tower onto an one of two other towers, shifting only one disc at a time, and making sure that a disc is never above a disc of smaller size. Thus, it might be said, to paraphrase and extend a quote by Neils Bohr, that ``if God exists, then not only does he play dice with the Universe, but he also plays Towers of Hanoi with the living creatures within it.'' 
It should be noted that there are other mathematical, biological, and mixed mathematical/biological reasons why the existing codes (both standard and the various mitochondrial and other codes) are optimal~\cite{GeneticCodeAdaptable} and also why there are differences between the various codes~\cite{MitochondrialCodes}.

\section{Game theory and cheating husbands}
\begin{figure}[hptb]
\centering
\includegraphics[width=4in]{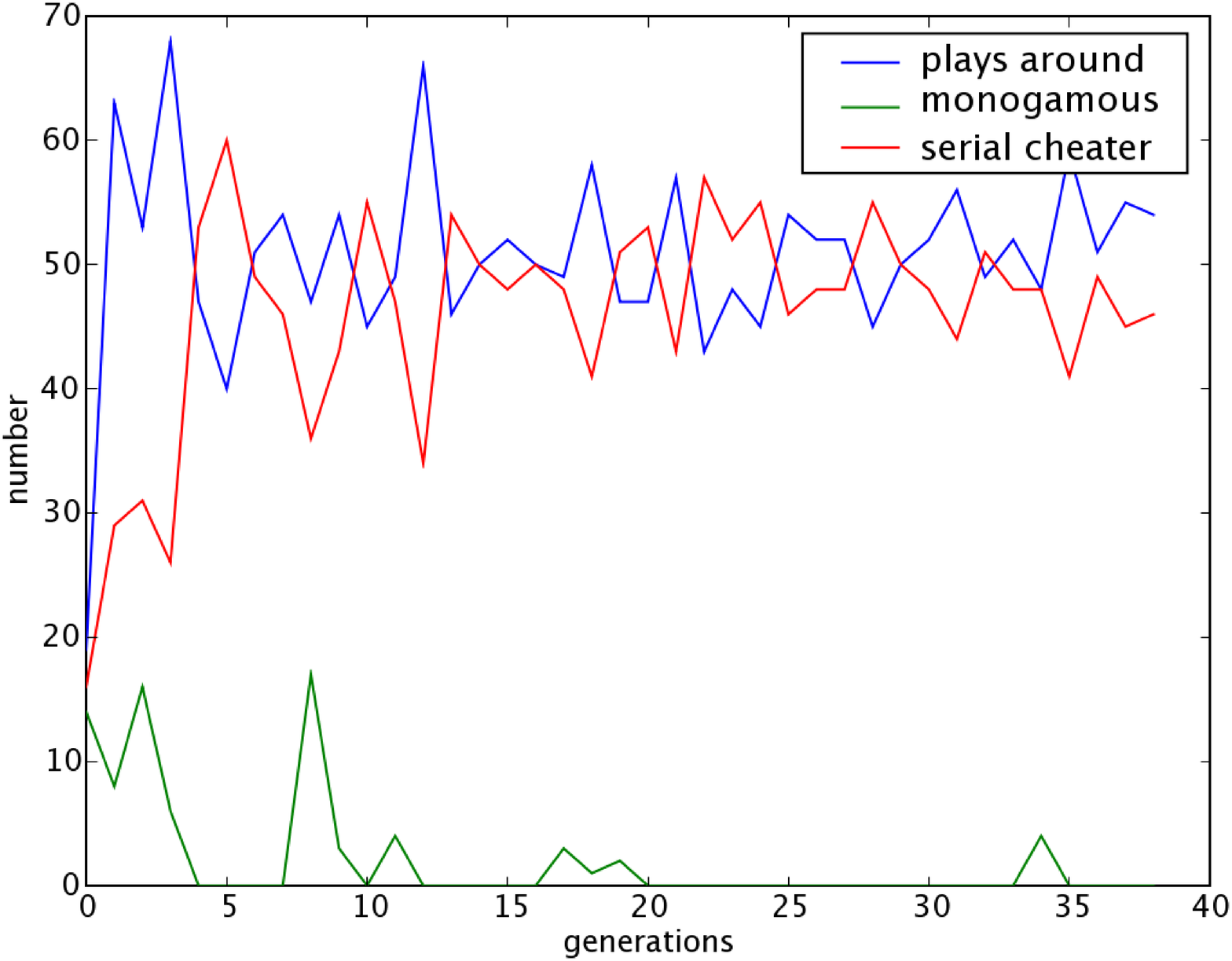}
\caption{Graph showing the population of people (1) playing around---mainly monogamous but with some cheating, (2) strictly monogamous, and (3) always playing around---never monogamous. Random fluctuations are present, due to the random aspects of the simulation. In general we obtain a population mix of people who are never monogamous or monogamous plus cheating. We expect this because there is an advantage to men to cheat, and neutral if women cheat. We don't expect a pure population of total cheaters because this would be penalised in general through the -2.0 score for women.}
\label{figure:fidelity_popn_graph}
\end{figure}

To answer the question of how genetic structures arise I considered a gene ``for'' infidelity. When I (and others, like Dawkins~\cite{SelfishGene} talk of a gene being ``for'' something, I am saying, all other things (including other genes) being equal, that this gene influences behaviour through different types (or chemical concentration) of a protein. It is also important to remember that along with genes, cultural and family background also play a large role in determining behaviour. The culture also sets up societal systems for taking care. Women face a number of tradeoffs in selecting mates, both for the long term and short term~\cite{WomenTradeOffs,EvolutionHumanMating} and these are highly dependent on culture, although it should be noted that some of these are also faced by females across the animal kingdom.
Some example scores (or relative advantages or disadvantages) for various male and female strategies are shown in Table~\ref{table:GTscores}. This assumes a society in which women are the predominant child raisers. If she can stay in a monogamous relationship, or ``get away'' with cheating and still have a husband around, then this is ``better'' ($1-(-2)>0$) than if she doesn't have a husband around. On the other hand, a male benefits his genes more and more, the more he is unfaithful, since the women will be raising his children. So it is not surprising that we therefore need a mixed strategy (which doesn't have to be used exclusively for a whole lifetime, although I simplified my simulation by doing that) in which there are some (women) who remain in semi-monogamous relationships, but there is a population of (mostly) men who cheat a lot. This is shown in Figure~\ref{figure:fidelity_popn_graph}, which I generated by performing a computer simulation capturing all of the above details. This whole model ignores emotions, but then in evolutionary terms, the emotions don't matter much here, since they mainly occur after children have already been raised. Thus we expect nature to not really care either way if people get hurt.

Of course, this is not the only possible system where game theory helps us determine evolutionary stable solutions to the problems that organisms face, this has also been shown for the threats from disease (both  genetic and infectious disease), we then find that these good solutions spread through a population: even if they confer a slight advantage, then over time they will spread. 
\begin{table}[hptb]
\centering
\caption{This table details the scores, or reproductive benefits to males and females who adopt various strategies: (1) playing around---mainly monogamous but with some cheating, (2) strictly monogamous, and (3) always playing around---never monogamous. We assume a society in which women are the predominant child rearers (found in some cultures but not others). Scores are only relative in determining overall success, for example, it wouldn't matter if we put -3 instead of -2, it is still less than 0 and so we expect that the $-3$ strategy will lose.}
\begin{tabular}{|c|c|c|}\hline
Gender & Fidelity & Score\\\hline
Male & Monogamous &  0\\
Male & Plays around & $\frac{1}{2}$\\
Male & Serial cheater & 1\\
Female & Monogamous & 1\\
Female & Plays around & 1\\
Female & Serial cheater & -2\\\hline
\end{tabular}
\label{table:GTscores}
\end{table}
\clearpage
\section{Entropy and introns}
In this section I will first introduce the topic of entropy, and then discuss how it applies to the introns, the parts of genes that are cut out of the transcribed mRNA sequence template before the protein is made. Entropy is also discussed later on, as it can also be used to analyse the mathematical properties of existing sequences.
\subsection{Entropy}
Entropy is a measure of the amount of order or disorder in a sequence, which can be thought of as the information (ignoring context). The mathematical formula is
\begin{equation}
H(\mathcal{X})=-\sum_{x\in \mathcal{X}}p_{x}\log_{2} p_{x},
\label{eqn:ShannonEntropy}
\end{equation}
where $x$ denotes different symbols from the set of symbols in a sequence, $\mathcal{X}$, and the $p_{x}$ is the probability of finding a symbol, or simply the number of times it occurs divided by the total number of symbols in the sequence. For example, the sequence $\mathcal{X}=aact$ has $p_{a}=2/4$, $p_{c}=p_{t}=1/4$, and thus has entropy $= -\left(2/4\log_{2}(2/4)+2\times1/4\log_{2}(1/4)\right)=1.5$ bits (the same bits that computers use) of information. A related topic to the Shannon entropy is Chaitin-Kolmogorov entropy. This is the ``algorithmic'' entropy, that is defined in terms of the shortest computer program that could reproduce a given sequence. This is related to the Shannon entropy (ideally it should approach, or get close to, the measure of the Shannon Entropy). We can consider the Chaitin-Kolmogorov entropy as being like a self-extracting ZIP (computer) file: the data is compressed, and a short program is attached which can then decompress the compressed data when the self-extracting file is run. I show below that this is similar to what occurs in introns

\subsection{Introns}
Entropy can enlighten us on two key things: evolutionary advantages for introns, and also on patterns found in specific existing genes. The former is discussed here, and the latter is discussed in the following section.

If we write consider each protein as composed of distinct functional modules (true for many, but not all proteins) then we often find other proteins containing the same modules. If we can write these alternative proteins as a single gene, with alternative splices, then we can increase the Shannon entropy, since there is less redundancy (and thus the probabilities of finding various bases are more even). This also increasess the Chaitin-Kolomogorov entropy, if we can use this alternative splicing a lot, in comparison to the extra genes we need to encode for this alternative splicing machinery---an ``algorithm'' to unpack the alternative splices from a single gene. In general, if the entropy of a system increases, the complexity increases (not always true since a true random signal has a very low complexity), and this leads to increased adaptability (but trades off reliability). 

The need to have minimal machinery here again guides us as to the evolutionary solution found. If we have some systematic way of marking where these modules, or exons, start and stop in genes, then we can use the same set of cellular machinery repeatedly. This then allows a greater degree of freedom in terms of the instructions that can be coded for, since we can include non-(protein-)coding instructions in these introns. As a very simple example of this, it has been showing that increasing the intron length can decrease the probability (or in other words, the final amount of protein) of containing the exon immediately after that intron. 

\section{Mathematical analysis of genetic structures}
Mathematics not only underpins genetic structures but it can also be used to analyse genetic structures in existing organisms. The following is an excerpt from my paper on using mutual information to analyse DNA sequences. Mutual information is like Shannon information above, except for two sequences. Basically it describes the total information covered by two sequences, say $\mathcal{X}$ and $\mathcal{Y}$, making sure to not double count the information they have in common. The mathematical formula is 
\begin{equation}
I(\mathcal{X}; \mathcal{Y}) = H(\mathcal{X})+H(\mathcal{Y})-H(\mathcal{X},\mathcal{Y}),
\label{eqn:mutualinfo}
\end{equation}
where $H(.)$ is the Shannon entropy defined above in Eq.~\ref{eqn:ShannonEntropy}

A mathematical for showing the existence of long-range correlations in DNA is to use the mutual information function, as given in Eq.~\ref{mutualinfo} below. This approach has been shown to distinguish between coding and non-coding regions~\cite{generatingcorr}.
We explore the use of the the mutual information function given in Eq.~\ref{mutualinfo}:
\begin{equation}
M(d)=\displaystyle \sum_{\alpha \in \mathit{A}} \sum_{\beta \in \mathit{A}} P_{\alpha \beta}(d) \log_{2}\frac{P_{\alpha \beta}(d)}{P_{\alpha}P_{\beta}},
\label{mutualinfo}
\end{equation}
for symbols $\alpha,\beta\in\mathit{A}$ (in the case of DNA, $\mathit{A}=\left\{a,t,c,g\right\}$). $P_{\alpha\beta}(d)$ is the probability that symbols $\alpha$ and $\beta$ are found a distance $d$ apart. This is related to the correlation function in Eq.~\ref{correlation}~\cite{mutualinfo}:
\begin{equation}
\Gamma (d) = \displaystyle  \sum_{\alpha \in \mathit{A}} \sum_{\beta \in \mathit{A}} a_{\alpha} a_{\beta}P_{\alpha\beta} (d)-\left(\sum_{\alpha \in \mathit{A}} a_{\alpha}P_{\alpha}\right)^{2},
\label{correlation}
\end{equation}
where $a_{\alpha}$ and $a_{\beta}$ are numerical representations of symbols $\alpha$ and $\beta$. As discussed by Li~\cite{mutualinfo}, the fact that we are working with a finite sequence means that this $M(d)$ overestimates the true $M_{\scriptscriptstyle T}(d)$ by
\begin{equation}
M(d)-M_{\scriptscriptstyle T}(d)\approx \frac{K\left(K-2\right)}{2N},
\label{overestimate}
\end{equation}
where $K$ is the number of symbols (for DNA this is always 4) and $N$ is the sequence length. 

\begin{figure}[hptb]
 \centering
\includegraphics[width=12cm]{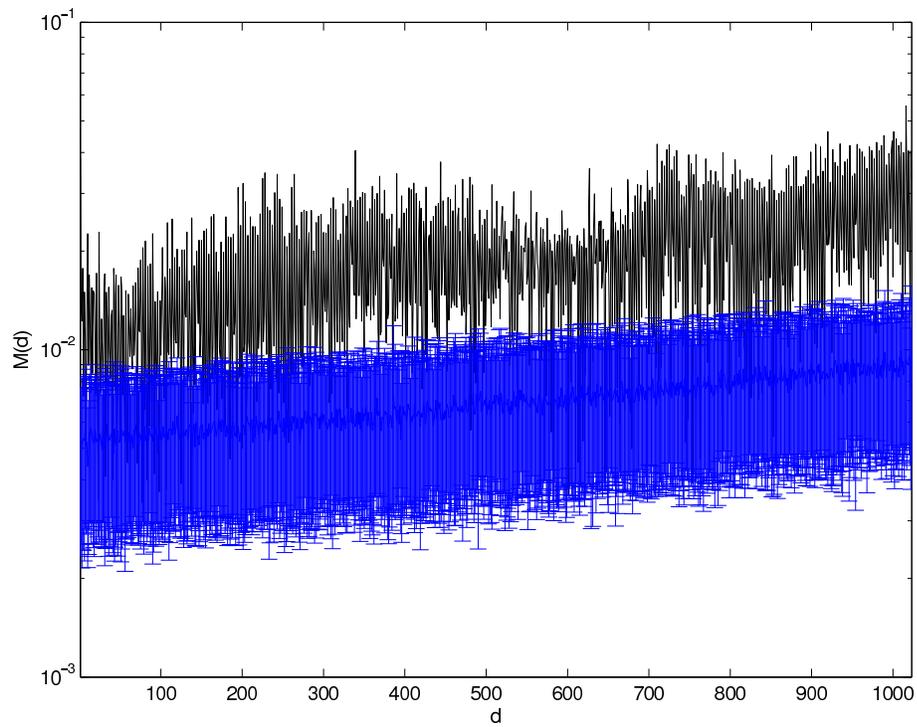}
\caption{This figure shows the plot of the mutual information function $M(d)$ in Eq.~\ref{mutualinfo} against base distance $d$ for the sequence of the MAP kinase-activated protein kinase 2 gene from {\it Mus musculus}  (in plain English: a mouse protein), shown in a darker line style, compared with the set of 100 randomized sequences of the same base distribution, the lighter band. The graph of mutual information in the MAP kinase gene mostly sits about the ``noise floor'' of the randomized sequences, in which the correlations have been destroyed}
\label{miplots:real}
\end{figure}

An example of applying this method to a real sequence of (mouse) DNA is shown in Figure~\ref{miplots:real}, clearly showing the existence of long-range correlations. It is not altogether clear why these correlations exist across proteins, it may be due to variants of functional modules, stringed together to make a protein, or it may be due to interesting structures in introns.
\section{Conclusions}
Mathematics presents us with powerful tools, such as Entropy, and Game Theory, that enlighten us as to what sort of genetic structures exist, how they evolve, and how we can analyse them.
In particular, I have shown mathematical arguments for:
\begin{itemize}
\item why four bases, a triplet code, and 20 amino acids are use,
\item why the triplets code for the 20 amino acids (and start and stop codons) in the way they do,
\item why introns are expected to evolve, and how they can be used to give increased flexibility,
\item how optimal solutions to evolutionary problems spread in a population, and
\item how to analyse genetic structures.
\end{itemize}

\end{document}